\documentclass[sigconf,nonacm]{acmart}
\AtBeginDocument{%
  \providecommand\BibTeX{{%
    \normalfont B\kern-0.5em{\scshape i\kern-0.25em b}\kern-0.8em\TeX}} }

\usepackage{changepage}
\usepackage{amsfonts}
\usepackage{algorithmic}
\usepackage{subfigure}
\usepackage{textcomp}
\usepackage{hyperref}
\usepackage{xcolor}
\usepackage{soul}
\usepackage{makecell}
\usepackage{multirow}
\usepackage{listings}
\definecolor{darkerGreen}{rgb}{0,0.7,0} % adjust values as needed
\setcopyright{none}
\begin{document}

\title[Analyzing Modern NVIDIA GPU cores]{Analyzing Modern NVIDIA GPU cores}
%\subtitle{\small ISCA 2025 Submission \textbf{\#821} – Confidential Draft – Do NOT Distribute!!}

\author{Rodrigo Huerta}
%\email{rodrigo.huerta.ganan@upc.edu}
\orcid{0000-0003-0052-7710}
\affiliation{%
  \institution{Universitat Politècnica de Catalunya}
  \city{Barcelona}
  \country{Spain}
}

\author{Mojtaba Abaie Shoushtary}
%\email{mojtaba.abaie@upc.edu}
\orcid{0000-0003-2377-6939}
\affiliation{%
  \institution{Universitat Politècnica de Catalunya}
  \city{Barcelona}
  \country{Spain}
}

\author{José-Lorenzo Cruz}
%\email{rodrigo.huerta.ganan@upc.edu}
\orcid{0000-0001-5325-9153}
\affiliation{%
  \institution{Universitat Politècnica de Catalunya}
  \city{Barcelona}
  \country{Spain}
}

\author{Antonio González}
%\email{antonio@ac.upc.edu}
\orcid{0000-0002-0009-0996}
\affiliation{%
  \institution{Universitat Politècnica de Catalunya}
  \city{Barcelona}
  \country{Spain}
}

\renewcommand{\shortauthors}{Huerta and Abaie, et al.}
%\renewcommand{\shortauthors}{Huerta et al.}

% Custom text variables
\newcommand{\bestStreamBufferSizeNumber}{16}
\newcommand{\numberOfGPUs}{4}
\newcommand{\numberOfGPUsWord}{four}
\newcommand{\numberOfBenchSuites}{12}
\newcommand{\numberOfDiffApplications}{83}
\newcommand{\numberOfDiffCombinations}{143}
\newcommand{\numberOfDiffKernels}{300} % calcular. No usado
\newcommand{\numberOfNotBinaryKernels}{6} % calcular. No usado
\newcommand{\targetArchitecture}{Ampere}
\newcommand{\targetGPU}{NVIDIA RTX A6000}
\newcommand{\targetGPUAmpere}{NVIDIA RTX A6000}
\newcommand{\targetGPUTuring}{NVIDIA RTX 2080 Ti }
\makeatletter
\newcommand*{\RESEVAL}[1]{%
  \stringcases
    {#1}%
    {%
      {MaeRTX2070SUs}{$10\%$}
      {MaeRTX2080TiUs}{$19.73\%$}
      {MaeRTX3080Us}{$17.15\%$}
      {MaeRTX3080TiUs}{$18\%$}
      {MaeRTX3090Us}{$17.93\%$}
      {MaeRTXA6000Us}{$13.98\%$}
      {MaeRTX2070SAccel}{$17\%$}
      {MaeRTX2080TiAccel}{$26.67\%$}
      {MaeRTX3080Accel}{$27.95\%$}
      {MaeRTX3080TiAccel}{$28.19\%$}
      {MaeRTX3090Accel}{$28.5\%$}
      {MaeRTXA6000Accel}{$32.22\%$}
      {CorrelRTX3080Us}{$0.99$}
      {CorrelRTX3080TiUs}{$0.99$}
      {CorrelRTX3090Us}{$0.98$}
      {CorrelRTXA6000Us}{$0.98$}
      {CorrelRTX2080TiUs}{$0.98$}
      {CorrelRTX3080Accel}{$0.98$}
      {CorrelRTX3080TiAccel}{$0.98$}
      {CorrelRTX3090Accel}{$0.98$}
      {CorrelRTXA6000Accel}{$0.97$}
      {CorrelRTX2080TiAccel}{$0.95$}
      {MaeBiggestImprovement}{$18.24\%$}
      {MaeBiggestImprovementTuring}{$6.94\%$}
      {MaeQ75Accel}{$31.45\%$} % calcular. No usado
      {MaeQ75Us}{$19.26\%$} % calcular. No usado
      {MaeQ90Accel}{$82.64\%$} % 
      {MaeQ90Us}{$31.47\%$}
      {AeBiggestErrorAccel}{$543\%$}
      {AeBiggestErrorUs}{$62\%$}
      {AeMoreThan100ErrorAccel}{10}
    }%
    {[MAE_ERROR_PLEASE_FIXME]}%
}
\newcommand{\stringcases}[3]{%
  \romannumeral
    \str@case{#1}#2{#1}{#3}\q@stop
}
\newcommand{\str@case}[3]{%
  \ifnum\pdf@strcmp{\unexpanded{#1}}{\unexpanded{#2}}=\z@
    \expandafter\@firstoftwo
  \else
    \expandafter\@secondoftwo
  \fi
    {\str@case@end{#3}}
    {\str@case{#1}}%
}
\newcommand{\str@case@end}{}
\long\def\str@case@end#1#2\q@stop{\z@#1}
\makeatother


\begin{abstract}
\par
GPUs are the most popular platform for accelerating HPC workloads, such as artificial intelligence and science simulations. However, most microarchitectural research in academia relies on GPU core pipeline designs based on architectures that are more than 15 years old.

\par
This paper reverse engineers modern NVIDIA GPU cores, unveiling many key aspects of its design and explaining how GPUs leverage hardware-compiler techniques where the compiler guides hardware during execution. In particular, it reveals how the issue logic works including the policy of the issue scheduler, the structure of the register file and its associated cache, and multiple features of the memory pipeline. Moreover, it analyses how a simple instruction prefetcher based on a stream buffer fits well with modern NVIDIA GPUs and is likely to be used. Furthermore, we investigate the impact of the register file cache and the number of register file read ports on both simulation accuracy and performance.

%\boldmath
\par
By modeling all these new discovered microarchitectural details, we achieve \RESEVAL{MaeBiggestImprovement} lower mean absolute percentage error (MAPE) in execution cycles than previous state-of-the-art simulators, resulting in an average of \RESEVAL{MaeRTXA6000Us} MAPE with respect to real hardware (\targetGPUAmpere). Also, we demonstrate that this new model stands for other NVIDIA architectures, such as Turing.
%\unboldmath

\par
Finally, we show that the software-based dependence management mechanism included in modern NVIDIA GPUs outperforms a hardware mechanism based on scoreboards in terms of performance and area.
\end{abstract}


% \ccsdesc[500]{Do Not Use This Code~Generate the Correct Terms for Your Paper\textnormal{; \textit{Generate
% the Correct Terms for Your Paper}; Generate the Correct Terms for
% Your Paper; Generate the Correct Terms for
% Your Paper}}

% %%
% %% Keywords. The author(s) should pick words that accurately describe
% %% the work being presented. Separate the keywords with commas.
% \keywords{GPU, Accel-sim, Dependence checking, Reverse engineering, Simulation, Modeling, Hardware software co-design, Instruction prefetching}


\maketitle

\section{Introduction}

\par
In recent years, GPUs have become very popular for executing general-purpose workloads~\cite{usageOfGPUs} in addition to graphics. GPUs' architecture provides massive parallelism that can be leveraged by many modern applications such as bioinformatics~\cite{cudaInGPUS, markovGPUBioinformatics}, physics~\cite{molecularPhysicsCuda, fdtdPhysicsCuda}, and chemistry~\cite{detailedChemistry, cheminformaticsGPU}, to name a few. Nowadays, GPUs are the main candidates to accelerate modern machine learning workloads, which have high memory bandwidth and computation demands~\cite{cuDNNLibraryIA}. Over the last years, there have been significant innovations in GPUs' microarchitecture, their interconnection technologies (NVLink~\cite{nvlink}), and their communication frameworks (NCCL~\cite{nccl}). All these advances have enabled inference and training of Large Language Models, which require clusters with thousands of GPUs~\cite{articuloWebMSNumGPUsIA}.

\par
However, there is scarce information on the microarchitecture design of modern commercial GPUs, and current academic studies~\cite{gpgpusim3, accelsim} take the Tesla microarchitecture~\cite{teslaHotchips} as the baseline, which was launched in 2006. Today GPU architectures has undergone significant changes since Tesla, hence, a model based on that can deviate the reported findings. This work aims to unveil different features and details of various components that modern NVIDIA GPU architectures use to improve the accuracy of academic microarchitecture models. The model and details explained in this work allow researchers to better identify challenges and opportunities for improving future GPUs. In summary, this paper makes the following contributions:

\begin{itemize}
    \item Describes the operation of the issue stage, including dependence handling, readiness conditions of warps, and the issue scheduler policy.
    \item Describes a plausible operation of the fetch stage and its scheduler that coordinates with the issue stage.
    \item Provides important details of the register file and explains the behavior of the register file cache. Moreover, it shows that modern NVIDIA GPUs do not use an operand collection stage or collector units.
    \item Reveals multiple details of the components of the memory pipeline.
    \item Redesigns the SM/core model used in Accel-sim simulator \cite{accelsim} from scratch and integrates all the details we revealed in this paper into the model.
    \item Validates the new model against real hardware and compares it against the Accel-sim simulator~\cite{accelsim}. Our new model achieves a mean absolute percentage error (MAPE) of execution cycles against real hardware of \RESEVAL{MaeRTXA6000Us} for the \targetGPUAmpere{} (Ampere), which is \RESEVAL{MaeBiggestImprovement} better than the previous simulator model.
    \item Demonstrates that a naive stream buffer for instruction prefetching provides a greater performance accuracy, and its performance is similar to a perfect instruction cache.
    \item Shows how the register file cache and the number of register file read ports affect simulation accuracy and performance.
    \item Compares the performance, area, and simulation accuracy of the dependence management system that we unveil in this paper against the traditional scoreboard. The comparison reveals that this novel software-hardware co-design is a more efficient alternative than handling dependencies with traditional scoreboards.
    \item Shows the portability of the model to other NVIDIA architectures such as Turing.
\end{itemize}

\par
The rest of this paper is organized as follows. First, we introduce the background and motivation of this work in \autoref{sec:backgroundmotivation}. In \autoref{sec:researchmethodology}, we explain the reverse engineering methodology that we have employed. We describe the control bits in modern NVIDIA GPU architectures and their detailed behavior in \autoref{sec:modernGPUArch}. Later, we present the core microarchitecture of these GPUs in \autoref{sec:design}. Next, we describe the features we have modeled in our simulator in \autoref{sec:modeling}. \hyperref[sec:valiadation]{Section~\ref{sec:valiadation}} evaluates the accuracy of our model against real hardware and compares it to the Accel-sim framework simulator, analyzes the impact of a stream buffer for instruction prefetching, studies the effect of the register file cache and the number of register file read ports, compares different dependence management mechanisms, and discusses how the model stands for other NVIDIA architectures. 
\hyperref[sec:relatedwork]{Section~\ref{sec:relatedwork}}  reviews previous related work. Finally, \autoref{sec:conclusion} summarizes the main lessons of this work.
\section{Background and Motivation}\label{sec:backgroundmotivation}

%Explicar algo de los CTA?
\par
\begin{figure}[ht]
  \centering
  \includegraphics[trim={0.1cm 0.1cm 0.1cm 0.1cm},clip,width=8.5cm]{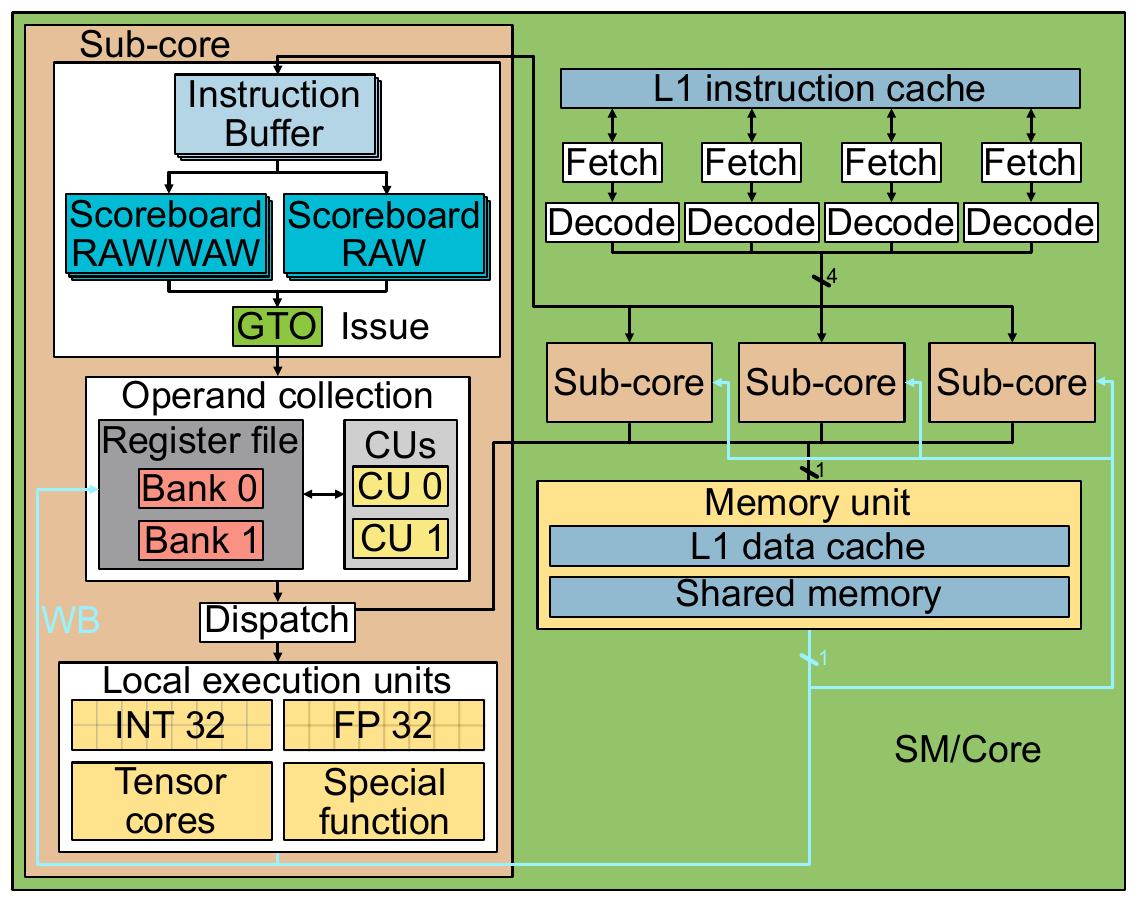}
  \caption{SM/Core academia design.}
  \label{fig:academia}
  \Description{Figure that shows the design of an SM/Core in academia} % Useless, used to remove warnings
\end{figure}

Most GPU microarchitecture research in academia relies on the microarchitecture that GPGPU-Sim simulator employs~\cite{gpgpusim3, gpgpuBook}. Recently, this simulator was updated to include sub-cores (Processing Blocks in NVIDIA terminology) approach that started in Volta.~\autoref{fig:academia} shows a block diagram of the architecture modeled in this simulator. We can observe that it comprises four sub-cores and some shared components, such as the L1 instruction cache, L1 data cache, shared memory, and texture units.

\par
In the Fetch stage of this GPU pipeline, a round robin scheduler selects a warp whose next instruction is in the L1 instruction cache and has empty slots in the Instruction Buffers. These buffers are dedicated per warp and store the consecutive instructions of a warp after they are fetched and decoded. Instructions stay in this buffer until they are ready and selected to be issued.

\par
In the Issue stage, a Greedy Then Oldest (GTO)~\cite{GTO} scheduler selects a warp if it is not waiting on a barrier and its oldest instruction does not have data dependence with other in-flight instructions in the pipeline. Previous works assume that each warp has two scoreboards for checking the data dependence. The first one marks pending writes to the registers to track WAW and RAW dependencies. An instruction can be issued only when all its operands are cleared in this scoreboard. The second scoreboard counts the number of in-flight consumers of registers to prevent WAR hazards~\cite{warHazards}. The second scoreboard is necessary because although instructions are issued in-order, their operands might be fetched out of order. This happens for variable-latency instructions such as memory instructions. These instructions are queued after being issued and may read its source operands after a younger arithmetic instruction writes its result, causing a WAR hazard if the source operand of the former is the same as the destination operand of the latter.

\par
Once an instruction is issued, it is placed in a Collector Unit (CU) and waits until all its source register operands are retrieved. Each sub-core has a private register file with multiple banks and a few ports per bank, allowing for multiple accesses in a single cycle at a low cost. An arbiter deals with the possible conflicts among several petitions to the same bank. When all source operands of an instruction are in the CU, the instruction moves to the Dispatch stage, where it is dispatched to the proper execution unit (e.g., memory, single-precision, special function) whose latencies differ depending on the unit type and instruction. Once the instruction reaches the write-back stage, the result is written in the register file.

\par
This GPU microarchitecture modeled in Accel-sim~\cite{accelsim} resembles NVIDIA GPUs based on Tesla~\cite{teslaHotchips}, which was released in 2006, and updated with a few modern features, mainly a sub-core model and sectored caches with IPOLY~\cite{IPOLYPaper} indexing similar to Volta~\cite{accelsim}. However, it lacks some important components that are present in modern NVIDIA GPUs, such as the L0 instruction cache~\cite{voltaPaper, turingPaper, amperePaper, hopperPaper, adaPaper, voltaHotChips, turingHotChips} and the uniform register file~\cite{turingHotChips}. Moreover, some main components of sub-cores, such as the issue logic, register file, or register file caches, among others, are not updated to reflect current designs.

\par
This work aims to reverse engineer the microarchitecture of the core in modern NVIDIA GPUs and update Accel-sim to incorporate the unveiled features. This will allow users of this updated Accel-sim simulator to make their work more relevant by starting with baselines closer to those proven successful by the industry in commercial designs.

\section{Reverse Engineering Methodology}\label{sec:researchmethodology}

\par
This section explains our research methodology for discovering the microarchitecture of the cores (SMs) in NVIDIA Ampere GPUs.

\par
Our approach is based on writing small microbenchmarks that consist of few instructions and measure the execution time of a particular small sequence of instructions. The elapsed cycles are obtained by surrounding a region of the code with instructions that save the \texttt{CLOCK} counter of the GPU into a register, and store it in main memory for later post-processing. The evaluated sequence of instructions typically consists of hand-written SASS instructions (and their control bits). Depending on the test, we visualize the recorded cycles to confirm or refute a particular hypothesis about the semantics of the control bits or a particular feature in the microarchitecture. Two examples to illustrate this methodology are given below:

\begin{itemize}
    \item We have used the code in \autoref{code:example_conflicts_rf} to unveil the conflicts of the multi-banked register file (\autoref{subsec:regfile}). Replacing \texttt{R\_X} and \texttt{R\_Y} by odd numbers (e.g., \texttt{R19} and \texttt{R21}), we get an elapsed time of five cycles (the minimum since each sub-core can issue one instruction per cycle). If we change \texttt{R\_X} to an even number (e.g., \texttt{R18}) while maintaining \texttt{R\_Y} odd (e.g., \texttt{R21}), the reported number of cycles is six. Finally, the elapsed time is seven cycles if we use an even number for both operands (e.g., \texttt{R18} and \texttt{R20}). In summary, two consecutive instructions can have from 0 to 2 cycles of bubbles in between, depending on which registers they use.
    \item \autoref{fig:issue_scheduling_timeline} is an example of a graphical representation of multiple recorded time marks, which in this case, it has been employed for discovering the issue policy of warps as explained in \hyperref[subsubsec:schedulingpolicy]{section 5.1.2}.
\end{itemize}

\begin{lstlisting}[caption={Code used to check Register File read conflicts.}, label={code:example_conflicts_rf}]
CLOCK
NOP
FFMA R11, R10, R12, R14
FFMA R13, R16, R_X, R_Y
NOP
CLOCK
\end{lstlisting}

\par
Although NVIDIA has no official tool to write SASS code (NVIDIA assembly language) directly, various third-party tools allow programmers to rearrange and modify assembly instructions (including control bits). These tools are used, for example, to optimize performance in critical kernels when the compiler generated code is not optimal. MaxAS~\cite{maxas} was the first tool for modifying SASS binaries. Later on, other tools such as KeplerAS~\cite{KeplerASRepo, KeplerASPaper} were developed for Kepler architecture. Then, TuringAS~\cite{turingas} and CUAssembler~\cite{CuAssembler} appeared to support more recent architectures. We have decided to use CUAssembler due to its flexibility, extensibility, and support for the latest hardware. 

\section{Control Bits in Modern NVIDIA GPU Architectures}\label{sec:modernGPUArch}

\par
The ISA of modern NVIDIA GPU architectures contains control bits and information that the compiler provides to maintain correctness. Unlike GPU architectures in previous works that check data dependence by tracking register reads and writes at run time (see~\autoref{sec:backgroundmotivation}), these GPU architectures rely on the compiler to handle register data dependencies~\cite{gtx680WhitePaper}. For this purpose, all assembly instructions include some control bits to properly manage dependencies in addition to improving performance and energy consumption.

\par
Below, we describe the behavior of these control bits included in every instruction. The explanation is based on some documents~\cite{dissectingVolta,dissectingTuring, CuAssembler, maxas}, but these documents are often ambiguous or incomplete, so we use the methodology described in~\autoref{sec:researchmethodology} to uncover the semantics of these control bits and verify that they act as described below.

\par
Sub-cores can issue a single instruction per cycle. By default, the Issue Scheduler tries to issue instructions of the same warp if the oldest instruction in the program order of that warp is ready. The compiler indicates when an instruction will be ready for issue using the control bits. If the oldest instruction of the warp that issued an instruction in the previous cycle is not ready, the issue logic selects an instruction from another warp, following the policy described in~\autoref{subsec:warpscheduler}.

\par
To handle producer-consumer dependencies of fixed-latency instructions, each warp has a counter that is referred to as \textit{\textbf{Stall counter}}. If this counter is not zero, this warp is not candidate to issue instructions. The compiler sets this counter with the latency of the producing instruction minus the number of instructions between the producer and the first consumer. All these per-warp \textit{Stall counters} are decreased by one every cycle until they reach 0. The issue logic will simply check this counter and will not consider issuing another instruction of the same warp until its value is zero.

\par
For example, an addition whose latency is four cycles and its first consumer is the following instruction encodes a four in the \textit{Stall counter}. Using the methodology explained in~\autoref{sec:researchmethodology}, we have verified that if the \textit{Stall counter} is not properly set, the result of the program is incorrect since the hardware does not check for RAW hazards, and simply relies on these compiler-set counters. In addition, this mechanism has benefits in terms of area and energy wiring. Keep in mind that wires from fixed-latency units to the dependence handling components are not needed, in contrast to a traditional scoreboard approach where they are required.

\par
Another control bit is called \textit{\textbf{Yield}}, and is used to indicate the hardware that in the next cycle it should not issue and instruction of the same warp. If the rest of the warps of the sub-core are not ready in the next cycle, no instruction is issued.

\par
Each instruction sets the \textit{Stall counter} and the \textit{Yield} bit. If the \textit{Stall counter} is greater than one, the warp will stall for at least one cycle, so in this case, no matter whether \textit{Yield} is set or not. 

% Si eso añadir en una version final

%However, experimentally we have observed that this is the case for \textit{Stall counters} up to the value of 11. On the other hand, if the \textit{Stall counter} is greater than 11 and the \textit{Yield} bit is not set, the warp stalls only 1 or 2 cycles. This is a very counter-intuitive combination and in fact, we never found this combination in the code generated by the compiler; we managed to test this case by modifying the control bits manually. We also found another special case, which is the \texttt{ERRBAR} instruction. This is the only instruction we have encountered that sets the \textit{Stall counter} to 0 and enables \textit{Yield}. We have measured that this combination makes the warp stall for 45 cycles before being allowed to issue another instruction of the same warp. 

\par
On the other hand, some instructions (e.g., memory, special functions) have variable latency, and the compiler does not know their execution time. Therefore, the compiler cannot handle these hazards through the \textit{Stall counter}. These hazards are resolved through \textit{\textbf{Dependence counters}} bits. Each warp has six special registers to store these counters, which are referred to as \texttt{SBx} with \texttt{x} taking a value in the range $[0-5]$. Each of these counters can count up to 63.

\par
These counters are initialized to zero when a warp starts. To handle a producer-consumer dependence, the producer increases a particular counter after issue, and decreases it at write-back. The consumer instruction is instructed to stall until this counter is zero.

\par
For WAR hazards, the mechanism is similar, with the only difference being that the counter is decreased after the instruction reads its source operands, instead of decreasing the counter at write-back.

\par
In each instruction, there are some control bits to indicate up to two counters that are increased at issue. One of these counters will be decreased at write-back (to handle RAW and WAW dependencies) and the other at register read (to handle WAR dependencies). For this purpose, every instruction has two fields of 3 bits each to indicate these two counters. Besides, every instruction has a mask of 6 bits to indicate which dependence counters it has to check to determine if it is ready for issue. Note that an instruction can check up to all six counters.

\par
Consider that if an instruction has multiple source operands whose producers have variable latency, the same \textit{Dependence counter} can be used by all these producers without losing any parallelism. It is important to note that this mechanism may encounter limits of parallelism in scenarios where there are more than six consumer instructions with different variable-latency producers. In such cases, the compiler must choose between two alternatives to manage the situation: 1) group more instructions under the same \textit{Dependence counter}, or 2) reorder the instructions differently.

\par
The incrementing of the \textit{Dependence counters} is performed the cycle after issuing the producer instruction, so it is not effective until one cycle later. Therefore, if the consumer is the next instruction, the producer has to set the \textit{Stall counter} to 2, to avoid issuing the consumer instruction the following cycle.

\begin{figure*}[ht]
  \centering
  \includegraphics[trim={0.1cm 0.1cm 0.1cm 0.1cm},clip,width=16cm]{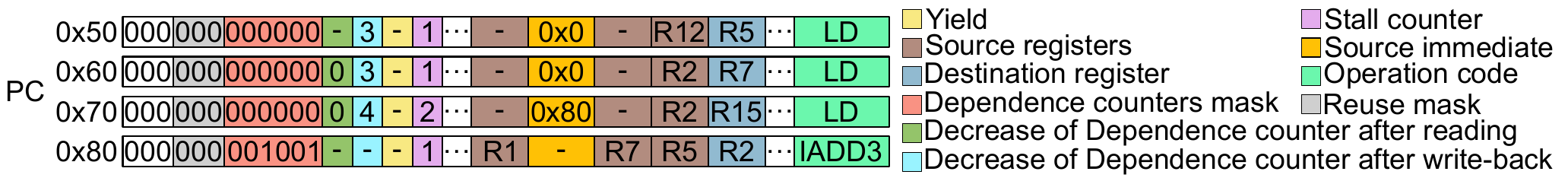}
  \caption{Example of using \textit{Dependence counters} to handle dependencies.}
  \label{fig:control_bits_example}
  \Description{Figure shows the behavior of Dependence counters to handle dependencies with multiple instructions} % Useless, used to remove warnings
\end{figure*}

\par
An example of handling dependencies with variable-latency producers can be found in~\autoref{fig:control_bits_example}. This code shows a sequence of four instructions (three loads and one addition) with its associated encoding. As the addition has dependencies with the loads (variable-latency instructions), \textit{Dependence counters} are used to prevent data hazards. The instruction at PC \texttt{0x80} has RAW dependencies with instructions at \texttt{0x50} and \texttt{0x60}. Thus, \texttt{SB3} is incremented by instructions \texttt{0x50} and \texttt{0x60} at issue and decremented at write-back. On the other hand, the addition has WAR dependencies with instructions at \texttt{0x60} and \texttt{0x70}. In consequence, \texttt{SB0} is increased by instructions \texttt{0x60} and \texttt{0x70} at issue and decreased after reading their respective register source operands. Finally, the \textit{\textbf{Dependence counters mask}} of the addition encodes that before being issued, \texttt{SB0} and \texttt{SB3} must be 0. Note that instruction  \texttt{0x70} also uses \texttt{SB4} to control RAW/WAR hazards with future instructions, but instruction \texttt{0x80} does not wait for this \textit{Dependence counter} since it does not have any dependence with that load. Clearing WAR dependencies after reading the source operands is an important optimization, since source operands are sometimes read much earlier than the result is produced, especially for memory instructions. For instance, in this example, instruction \texttt{0x80} waits until instruction \texttt{0x70} reads \texttt{R2} for clearing this WAR dependence, instead of waiting until instruction \texttt{0x70} performs its write-back, which may happen hundreds of cycles later.

\par
An alternative way of checking the readiness of these counters is through the \texttt{DEPBAR.LE} instruction. As an example, \texttt{DEPBAR.LE SB1, 0x3, \{4,3,2\}}, requires the \textit{Dependence counter} \texttt{SB1} to have a value less or equal to 3 to continue with the execution. The last argument (\texttt{[, \{4,3,2\}]}) is optional, and if used, the instruction cannot be issued until the values of the \textit{Dependence counters} specified by those IDs (4, 3, 2 in this example) are equal to 0.

\par
\texttt{DEPBAR.LE} can be especially useful in some particular scenarios. For instance, it allows the use of the same \textit{Dependence counter} for a sequence of $N$ variable-latency instructions that perform their write-back in order (e.g., memory instructions with the \texttt{STRONG.SM} modifier) when a consumer needs to wait for the first $M$ instructions. Using a \texttt{DEPBAR.LE}  with its argument equal to $N-M$ makes this instruction wait for the $M$ first instructions of the sequence. Another example is to reuse the same \textit{Dependence counter} to protect RAW/WAW and WAR hazards. If an instruction uses the same \textit{Dependence counter} for both types of hazards, as WAR hazards are resolved earlier than RAW/WAW, a following \texttt{DEPBAR.LE SBx, 0x1} will wait until the WAR is solved and allow the warp to continue its execution. A later instruction that consumes its result needs to wait until this \textit{Dependence counter} becomes zero, which means that the results have been written.

\par
Additionally, GPUs have a register file cache used to save energy and reduce contention in the register file read ports. This structure is software-managed by adding a control bit to each source operand, the \textit{\textbf{reuse}} bit,  which indicates the hardware whether to cache or not the content of the register. More details about the organization of register file cache are explained in \hyperref[subsubsec:reg_file_cache]{section 5.3.1}. %\autoref{subsubsec:reg_file_cache}.

\par
Finally, although this paper focuses on NVIDIA architectures, exploring AMD GPU ISAs documentation~\cite{amdCDNA1,amdCDNA2,amdCDNA3,amdRDNA1,amdRDNA2,amdRDNA35,amdRDNA3,amdVega7nm,amdGCN3,amdVega} reveals that AMD also relies on a hardware-software codesign to manage dependencies and boost performance. Similar to NVIDIA’s \texttt{DEPBAR.LE} instruction, AMD employs a \texttt{waitcnt} instruction; depending on the architecture, each wavefront (warp) has three or four counters, with each counter dedicated to specific instruction types and its use required to protect data hazards created by those instructions. AMD does not allow regular instructions to wait for a counter to reach zero using control bits, requiring an explicit \texttt{waitcnt} instruction instead, which increases the number of instructions. This design reduces the decoding overhead, yet increases the overall instruction count. In contrast, NVIDIA’s alternative enables more concurrent dependence chains even within the same instruction type, as it has up to two counters more per warp, and the counters are not tied to any instruction type. Although AMD does not need software or compiler intervention to avoid data hazards with ALU instructions, it introduced the \texttt{DELAY\_ALU} instruction in RDNA 3/3.5 architectures~\cite{amdRDNA35, amdRDNA3} to mitigate pipeline stalls caused by dependencies. Conversely, NVIDIA depends on the compiler to correctly handle data dependencies by setting the \textit{Stall counter} for fixed-latency producers, resulting in a lower instruction count but higher decoding overhead.
\section{GPU cores Microarchitecture}\label{sec:design}

\par
\begin{figure}[ht]
  \centering
  \includegraphics[trim={0.1cm 0.1cm 0.1cm 0.1cm},clip,width=8.5cm]{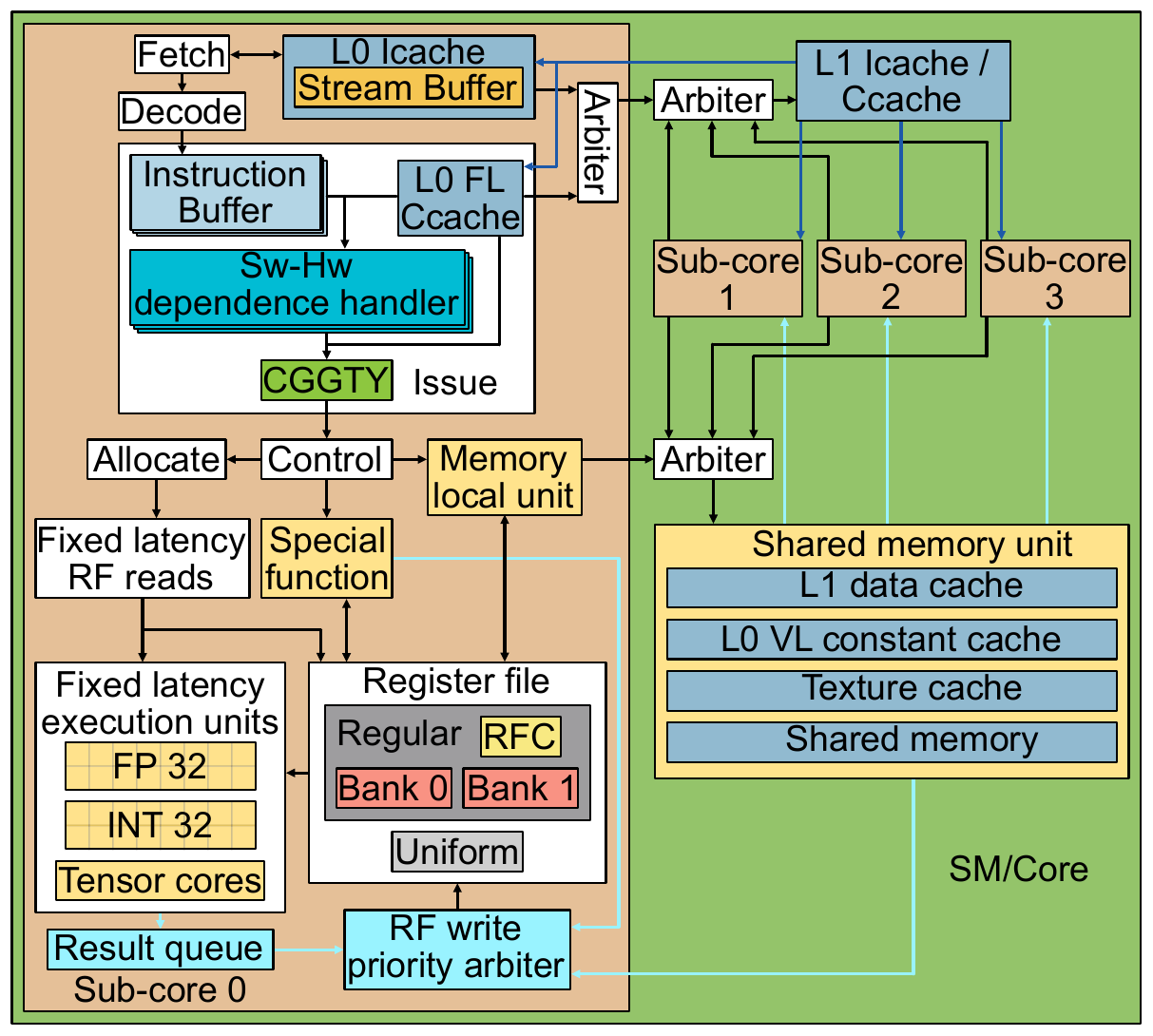}
  \caption{Modern NVIDIA GPU SM/Core design.}
  \Description{Figure that shows the design of our model for the SM/Core} % Useless, used to remove warnings
  \label{fig:ourCore}
  \vskip -0.3cm
\end{figure}

In this section, we describe our findings regarding the microarchitecture of GPU cores of modern commercial NVIDIA GPUs, using the methodology explained in \autoref{sec:researchmethodology}. \autoref{fig:ourCore} shows the main components of the GPU cores' microarchitecture. Below, we describe in detail the microarchitecture of the issue scheduler, the front-end, the register file, and the memory pipeline.

\subsection{Issue Scheduler}\label{subsec:warpscheduler}

\par
In this subsection we dissect the Issue Scheduler of modern NVIDIA GPUs. First, we describe which warps are considered candidates for issue every cycle in \autoref{subsubsec:readinesschecking}. Then, we present the Selection policy in \autoref{subsubsec:schedulingpolicy}. 

\begin{figure*}[t!]
    \centering
    \includegraphics[width=\textwidth]{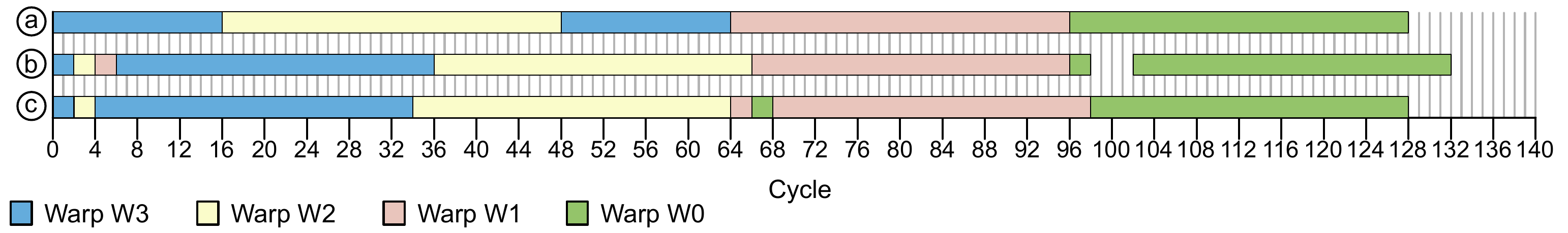}
    \caption{Timelines of the issue of instructions from four different warps.}
    \label{fig:issue_scheduling_timeline}
    \Description{Figure with three different timelines. Each of them, shows the issue slots of four different warps} % Useless, used to remove warnings
\end{figure*}

\subsubsection{Warp Readiness}\label{subsubsec:readinesschecking}

\par
A warp is considered a candidate for issuing its oldest instruction in a given cycle if some conditions are met. Some of these conditions depend on previous instructions of the same warp, while others rely on the global state of the core.

\par
An obvious condition is having a valid instruction in the Instruction Buffer. Another condition is that the oldest instruction of the warp must not have any data dependence hazard with older instructions of the same warp not yet completed. Dependencies among instructions are handled through software support by means of the control bits described above in \autoref{sec:modernGPUArch}. 

\par
Besides, for fixed-latency instructions, a warp is candidate to issue its oldest instruction in a given cycle only if it can be guaranteed that all needed resources for its execution will be available once issued. 

\par
One of these resources is the execution unit. Execution units have an input latch that must be free when the instruction reaches the execution stage. This latch is occupied for two cycles if the width of the execution unit is just half warp, and for one cycle if its width is a full warp. 

\par
For instructions that have a source operand in the Constant Cache, the tag look-up is performed in the issue stage. When the oldest instruction of the selected warp requires an operand from the constant cache and the operand is not in the cache, the scheduler does not issue any instruction until the miss is serviced. However, if the miss has not been served after four cycles, then the scheduler switches to a different warp (the youngest with a ready instruction). 

%That is, once these instructions are issued they are never stalled in any pipeline stage. 
\par
As for the availability of read ports in the register file, the issue scheduler is unaware of whether the evaluated instructions have enough ports for reading without stalls in the following cycles. We have reached this conclusion after observing that the conflicts in the code shown in \autoref{code:example_conflicts_rf} do not stall the issue of the second \texttt{CLOCK} if we remove the \texttt{NOP} between the last \texttt{FFMA} and the last \texttt{CLOCK} instruction. We performed a multitude of experiments to unveil the pipeline structure between issue and execute, and we could not find a model that perfectly fits all the experiments. However, the model that we describe below is correct for almost all the cases, so it is the model that we assume. In this model, fixed-latency instructions have two intermediate stages between the Issue stage and the stage(s) for reading source operands. The first stage, which we call Control, is common for fixed and variable latency instructions, and its duty is to increase the \textit{Dependence counters} or read the value of the clock counter if needed. This causes, as corroborated by our experiments, that an instruction that increases the \textit{Dependence counter} and the instruction that waits until that \textit{Dependence counter} is 0 requires at least one cycle in between to make visible that increase, so two consecutive instructions cannot use \textit{Dependence counters} to avoid data dependence hazards unless the first instruction sets the \textit{Yield} bit or a \textit{Stall counter} bigger than one.

\par
The second stage only exists for fixed-latency instructions. In this stage, the availability of register file read ports is checked, and the instruction is stalled in this stage until it is guaranteed that it can proceed without any register file port conflict. We call this stage Allocate. More details about the register file read and write pipeline and its cache are provided in \autoref{subsec:regfile}.

\par
Variable-latency instructions (e.g., memory instructions) are delivered directly to a queue after going through the Control stage (without going through the Allocate stage). Instructions in this queue are allowed to proceed to the register file read pipeline when they are guaranteed not to have any conflict. Fixed-latency instructions are given priority over variable-latency instructions to allocate register file ports, as they need to be completed in a fixed number of cycles after issue to guarantee the correctness of the code, since dependencies are handled by software as described above.

\subsubsection{Scheduling Policy}\label{subsubsec:schedulingpolicy}

\par
To discover the policy of the issue scheduler, we developed many different test cases involving multiple warps and recorded which warp was chosen for issue in each cycle by the issue scheduler. This information was gathered through instructions that allow to save the current \texttt{CLOCK} cycle of the GPU. However, as the hardware does not allow issuing two of these instructions consecutively, we employed a controlled number of other instructions in between (normally \texttt{NOP}s). We also varied the specific values in the \textit{Yield} and the \textit{Stall counter} control bits. 

\par
Our experiments allowed us to conclude that the warp scheduler uses a greedy policy that selects an instruction from the same warp if it meets the eligibility criteria described above. When switching to a different warp, the youngest one that meets the eligibility criteria is selected. 

\par
This issue scheduler policy is illustrated with some examples of our experiments in \autoref{fig:issue_scheduling_timeline}. This figure depicts the issue of instructions when four warps are executed in the same sub-core for three different cases. Each warp executes the same code composed of 32 independent instructions that can be issued one per cycle.

\par
In the first case, \autoref{fig:issue_scheduling_timeline} {\large \textcircled{\normalsize a}}, all \textit{Stall counters}, \textit{Dependence masks} and \textit{Yield} bits are set to zero. The scheduler starts issuing instructions from the youngest warp, which is W3, until it misses in the Icache. As a result of the miss, W3 does not have any valid instruction, so the scheduler switches to issue instructions from W2. W2 hits in the Icache since it reuses the instructions brought by W3, and when it reaches the point where W3 missed, the miss has already been served, and all remaining instructions are found in the Icache, so the scheduler greedily issues that warp until the end. Later, the scheduler proceeds to issue instruction from W3 (the youngest warp) until the end, since now all instructions are present in the Icache. Then, the scheduler switches to issue instructions from W1 from beginning to end, and finally, it does the same for W0 (the oldest warp).

\par
\autoref{fig:issue_scheduling_timeline} {\large \textcircled{\normalsize b}} shows the timeline of when instructions are issued when the second instruction of each warp sets its \textit{Stall counter} to four. We can observe that the scheduler swaps from W3 to W2 after two cycles, to W1 after another two cycles, and then back to W3 after another two cycles (since W3 \textit{Stall counter} has become zero). Once W3, W2, and W1 have finished, the scheduler starts issuing from W0. After issuing the second instruction of W0, the scheduler generates four bubbles because it has no other warp to hide the latency imposed by the \textit{Stall counter}. 

\par
\autoref{fig:issue_scheduling_timeline} {\large \textcircled{\normalsize c}} shows the scheduler's behavior when \textit{Yield} is set in the second instruction of each warp. We can see that the scheduler switches to the youngest among the rest of the warps after issuing the second instruction of each warp. For instance, W3 switches to W2, and W2 switches back to W3. We also tested a scenario where \textit{Yield} is set and no more warps are available (not shown in this figure), and we observed that the scheduler generates a bubble of one cycle.

\par
We call this issue scheduler policy Compiler Guided Greedy Then Youngest  (CGGTY) since the compiler assists the scheduler by means of the control bits: \textit{Stall counter}, \textit{Yield} and \textit{Depencence counters}.

\par
However, we have only confirmed this behavior for warps within the same CTA, as we have not yet devised a reliable methodology to analyze interactions among warps from different CTAs.

\subsection{Front-end}\label{subsubsec:frontend}

\par
According to diagrams in multiple NVIDIA documents~\cite{voltaPaper, turingPaper, amperePaper, hopperPaper, adaPaper, voltaHotChips, turingHotChips}, SMs have four different sub-cores and warps are evenly distributed among sub-cores in a round robin manner (i.e., warp ID $\% 4$)~\cite{dissectingVolta, dissectingTuring}. Each sub-core has a private L0 instruction cache that is connected to an L1 instruction cache that is shared among all four sub-cores of the SM. We assume there is an arbiter for dealing with the multiple requests of different sub-cores. 

%However, the peak of the L1 instruction cache serving four instructions per clock is still possible if the four sub-cores request the same instruction. 

\par
Each L0 Icache has an instruction prefetcher \cite{nvidiaInstPrefeching}. Our experiments corroborate previous studies by Cao et al. \cite{GPUinstPrefeching} that demonstrated that instruction prefetching is effective in GPUs. Although we have not been able to confirm the concrete design used in NVIDIA GPUs, we suspect it is a simple scheme like a stream buffer \cite{streamBufferPrefetching} that prefetches successive memory blocks when a miss occurs. We assume that the stream buffer size is \bestStreamBufferSizeNumber{} based on our analysis, as detailed in \autoref{subsec:results_instruction_prefetching}. 

\par
We could not confirm the exact instruction fetch policy with our experiments, but it has to be similar to the issue policy; otherwise, the condition of not finding a valid instruction in the Instruction Buffer would happen relatively often, and we have not observed this in our experiments. Based on that, we assume that each sub-core can fetch and decode one instruction per cycle. The fetch scheduler tries to fetch an instruction from the same warp that has been issued in the previous cycle (or the latest cycle in which an instruction was issued) unless it detects that the number of instructions already in the Instruction Buffer plus its in-flight fetches are equal to the Instruction Buffer size. In this case, it switches to the youngest warp with free entries in its Instruction Buffer. We assume an Instruction Buffer with three entries per warp since this is enough to support the greedy nature given that there are two pipeline stages from fetch to issue. Were the Instruction Buffer of size two, the Greedy policy of the issue scheduler would fail. For example, assume a scenario in which the Instruction Buffer has a size of two and all the requests hit in the Icache, all warps have their Instruction Buffer full, and in cycle 1, a sub-core is issuing instructions from warp W1 and fetching from W0. In cycle 2, the second instruction of W1 will be issued, and the third instruction will be fetched. In cycle 3, W1 will have no instructions in its Instruction Buffer because instruction 3 is still in the decode. Therefore, its greedy behavior would fail, and it would have to switch to issue from another warp. Note that this would not happen in case of having three entries in the Instruction Buffer, as corroborated by our experiments. Note that most previous designs in the literature normally assume a fetch and decode width of two instructions and an Instruction Buffer of two entries per warp. In addition, those designs only fetch instructions when the Instruction Buffer is empty. Thus, the greedy warp always changes at least after two consecutive instructions, which does not match our experimental observations.  

%\par
%We embrace the idea of having a private instruction Buffer per warp with three entries to be more literature-compliant. However, other designs can be functionally equivalent.

\subsection{Register File}\label{subsec:regfile}

\par
We have performed a multitude of experiments by running different combinations of SASS assembly instructions to unveil the register file organization. For example, we wrote codes with different pressure on the register file ports, with and without using the register file cache.

\par
Modern NVIDIA GPUs have various register files:

\begin{itemize}
    \item \textbf{Regular:} Recent NVIDIA architectures have 65536 32-bit registers per SM~\cite{voltaPaper, turingPaper, amperePaper, hopperPaper, adaPaper} used to store the values operated by threads. The registers are arranged in groups of 32, each group corresponding to the registers of the 32 threads in a warp, resulting in 2048 warp registers. These registers are evenly distributed between sub-cores, and the registers in each sub-core are organized in two banks~\cite{dissectingVolta, dissectingTuring}. The number of registers used by a particular warp can vary from 1 to 256, and it is decided at compile time. The more registers used per warp, the fewer warps can run parallel in the SM.
    \item \textbf{Uniform:} Each warp has 64 private, 32-bit registers that store values shared by all the threads of the warp~\cite{dissectingTuring}.
    \item \textbf{Predicate:} Each warp has eight 32-bit registers, each bit being used by a different thread of the warp. These predicates are used by warp instructions to indicate which threads must execute the instruction and, in the case of branches, which threads must take the branch and which ones not.
    \item \textbf{Uniform Predicate:} Each warp has eight 1-bit registers that store a predicate shared by all the treads in the warp.
    \item \textbf{SB Registers:} As described in~\autoref{sec:modernGPUArch}, each warp has six registers, called \textit{Dependence counters}, that are used to track variable-latency dependencies. 
    \item \textbf{B Registers:} Each warp has at least 16 \texttt{B} registers for managing control flow re-convergence~\cite{mojtabaControlFlow}.
    \item \textbf{Special Registers:} Various other registers are used to store special values, such as the thread or block IDs.
\end{itemize}

\par
Unlike previous works~\cite{GPU2OcusSubcore, accelsim, gpgpuBook} that assume the presence of an operand collector to deal with conflicts in the register file ports, modern NVIDIA GPUs do not make use of it. Operand collector units would introduce variability in the elapsed time between issue and write-back, making it impossible to have fixed-latency instructions for the NVIDIA ISA, whose latency must be known at compile time to handle dependencies correctly as explained~\autoref{sec:modernGPUArch}. We have confirmed the absence of operand collectors by checking the correctness of specific producer-consumer sequences of instructions when varying the number of operands that go to the same bank. We observed that regardless of the number of register file port conflicts, the value required in the \textit{Stall counter} field of instructions to avoid data hazards and the elapsed time to execute the instruction remains constant.

\par
Our experiments revealed that each register file bank has a dedicated write port of 1024 bits. Besides, when a load instruction and a fixed-latency instruction finish at the same cycle, the one that is delayed one cycle is the load instruction. On the other hand, when there is a conflict between two fixed latency instructions, for instance, a \texttt{IADD3} followed by an \texttt{IMAD} that uses the same destination bank, none of them is delayed. This implies the use of a result queue like the one introduced in Fermi \cite{fermiWhitePaper} for fixed-latency instructions. The consumers of these instructions are not delayed, which implies the use of bypassing to forward the results to the consumer before being written in the register file.

\par
Regarding reads, we have observed a bandwidth of 1024 bits per bank. The measurements were obtained through various tests that recorded the elapsed time of consecutive \texttt{FADD}, \texttt{FMUL}, and \texttt{FFMA} instructions \footnote{Ampere allows the execution of FP32 operations into the FP32 and INT32 execution units \cite{amperePaper}. Therefore, there are no bubbles between two consecutive FP32 instructions due to execution unit conflicts.}. For instance, \texttt{FMUL} instructions with both source operands in the same bank create a 1-cycle bubble, whereas if the two operands are in different banks, there is no bubble. \texttt{FFMA} instructions with all three source operands in the same bank generate a 2-cycle bubble. 

\par
Unfortunately, we could not find a read policy that matches all the cases we have studied, as we observed that the generation of bubbles depends on the type of instructions and the task of each operand in the instructions. We found that the best approximation that matches almost all the experiments we tested is a scheme with two intermediate stages between the instruction issue and operand read of fixed-latency instructions, which we call Control and Allocate. The former has been explained in \autoref{subsubsec:readinesschecking}. The latter is in charge of reserving the register file read ports. Each bank of the register file has one read port of 1024 bits, and read conflicts are alleviated by using a register file cache (more details later). Our experiments showed that all the fixed-latency instructions spend three cycles for reading source operands, even if some of these cycles the instruction is idle (for instance when there are only two source operands) because \texttt{FADD} and \texttt{FMUL} have the same latency as the \texttt{FFMA} despite having one operand less, and \texttt{FFMA} always has the same latency regardless of whether its three operands are in the same bank or not. If the instruction in the allocate stage realizes that it cannot read all its operands in the next three cycles, it is held in this stage (stalling the pipeline upwards) and generates bubbles until it can reserve all ports needed to read the source operands in the next three cycles. 

%Deberiamos de decir algo de bypasses?

%\par
%As some instructions use more than one 32-bit register per source operand per thread (1024-bit per warp), this introduces constraints on what registers can be encoded in specific instructions. For example, memory references to global memory addresses (64 bits) require two 32-bit registers for each address. The source operand is specified as an even number register implying this register and the next one (e.g., \texttt{R6} and \texttt{R7}).

\subsubsection{Register file cache}\label{subsubsec:reg_file_cache}

% \begin{figure}[ht]
%   \centering
%   \includegraphics[trim={0.5cm 0.5cm 0.5cm 0.5cm},clip,width=8cm]{figures/Rev_RegisterFile.pdf}
%   \caption{Regular Register File.}
%   \label{fig:registerFile}
% \end{figure}

% \autoref{fig:registerFile} shows a block diagram of the regular register file. 

\par
The use of a Register file cache (RFC) in GPUs has been investigated for relieving contention in the Register File ports and saving energy~\cite{registerFileCacheFirst, registerFileCacheSimilar, BOW, ltrf, mojtabaIsca}. 

\par
Through our experiments, we observed that the NVIDIA design is similar to the work of Gebhart et al. \cite{registerFileCacheSimilar}. In line with that design, the RFC is controlled by the compiler and is only used by instructions that have operands in the Regular Register File. Regarding the Last Result File structure, what we call the result queue behaves similarly. However, unlike the above-cited paper, a two-level issue scheduler is not used, as explained above in \autoref{subsubsec:schedulingpolicy}. 

\par
Regarding the organization of the RFC, our experiments showed that it has one entry for each of the two register file banks in each sub-core. Each entry stores three 1024-bit values, each corresponding to one of the three regular register source operands that instructions may have. Overall, the RFC's total capacity is six 1024-bit operand values (sub-entries). Note that there are instructions that have some operands that require two consecutive registers (e.g., tensor core instructions). In this case, each of these two registers come from a different bank, and are cached in their corresponding entries.

\par
The compiler manages the allocation policy. When an instruction is issued and reads its operands, each operand is stored in the RFC if the compiler has set its reuse bit for that operand. A subsequent instruction will obtain its register source operand from the RFC if the instruction is from the same warp, the register ID coincides with the one stored in the RFC, and the operand position in the instruction is the same as in the instruction that has triggered the caching. A cached value is unavailable after a read request arrives to the same bank and operand position, regardless of whether it hits in the RFC. This is illustrated in example 2 of \autoref{code:rfc}; to allow the third instruction to find \texttt{R2} in the cache, the second instruction must set the reuse bit of \texttt{R2} in spite of \texttt{R2} being already in the cache for the second instruction. \autoref{code:rfc} shows three other examples to illustrate the RFC behavior.

\begin{lstlisting}[caption={Register file cache behavior.}, label={code:rfc}]
# Example 1
IADD3 R1, R2.reuse, R3, R4 # Allocates R2
FFMA R5, R2, R7, R8 # R2 hits and becomes unavailable
IADD3 R10, R2, R12, R13 # R2 misses

# Example 2
IADD3 R1, R2.reuse, R3, R4 # Allocates R2
FFMA R5, R2.reuse, R7, R8 # R2 hits and is retained
IADD3 R10, R2, R12, R13 # R2 hits

# Example 3. R2 misses in the second instruction since it is 
# cached in another slot. R2 remains available in the first
# slot since R7 uses a different bank
IADD3 R1, R2.reuse, R3, R4 # Allocates R2
FFMA R5, R7, R2, R8 # R2 misses
IADD3 R10, R2, R12, R13 # R2 hits 

# Example 4. R2 misses in the third instruction since the 
# second instruction uses a different register that goes to
# the same bank in the same operand slot
IADD3 R1, R2.reuse, R3, R4 # Allocates R2
FFMA R5, R4, R7, R8 # R4 misses and R2 becomes unavailable
IADD3 R10, R2, R12, R13 # R2 misses

\end{lstlisting}

\subsection{Memory Pipeline}\label{subsec:memory_pipeline}

\par
The memory pipeline in modern NVIDIA GPUs has some initial stages local to each sub-core, whereas the last stages that perform the memory access are shared by the four sub-cores since the data cache and the shared memory are shared by them \cite{voltaHotChips, turingHotChips}. In this section, we discover the size of the load/store queues in each sub-core, the rate at which each sub-core can send requests to the shared memory structures, and the latency of the different memory instructions. 

\par
Note that there are two main types of memory accesses, those that go to shared memory (the SM local memory shared among all threads in a block) and those that go to the global memory (the GPU main memory).

\par
To discover the size of the queues and the memory bandwidth, we run a set of experiments in which each sub-core either executes a warp or is idle. Each warp executes a sequence of independent loads or stores that always hit in the data cache or shared memory and use regular registers. \autoref{tab:memoryconsecutive} shows the results of these experiments. The first column shows the instruction number in the code, and the next four columns show in which cycle this instruction is issued in each of the cores for four different scenarios that differ in the number of active cores.

\par
We can observe that each sub-core can issue one instruction per cycle for five consecutive memory instructions in Ampere. The issue of the 6th memory instruction is stalled for a number of cycles that depend on the number of active sub-cores. From these data, we can infer that each sub-core can buffer up to five consecutive instructions without stalling, and the global structures can receive a memory request every two cycles from any of the sub-cores. For instance, if we look at the cases with multiple active sub-cores, we can see that the 6th and following instructions of each sub-core are issued every other two cycles.

\par
We can also infer that the address calculation done in each sub-core has a throughput of one instruction every four cycles, as demonstrated by the 4-cycle gap in the issue after the 6th instruction when there is a single active sub-core. When two sub-cores are active, each can issue a memory instruction every 4 cycles since the shared-structures can handle one instruction every two cycles. When more sub-cores are active, the shared-structures become the bottleneck. For instance, when four sub-cores are active, each sub-core can issue an instruction only every 8 cycles since the shared-structures only allow one instruction every two cycles as its maximum throughput.

\par
Regarding the size of the memory queue in each sub-core, we estimate that it has a size of four even though each sub-core can buffer five consecutive instructions. The instruction reserves the slot when it arrives at the unit and frees it when it leaves the unit. 

%Depending on the instruction, it goes through more or less phases in the sub-core and spends a different amount of cycles on each of these phases.

\begin{table}
    \centering
    \small
    \begin{tabular}{|c|c c c c|}
    \hline
    \multirow{2}{*}{\shortstack{Instruction\\\#}} & \multicolumn{4}{c|}{\# Active sub-cores} \\
%    \cline{2-5}
     & 1 & 2 & 3 & 4 \\
    \hline
    1 & 2 & 2/2 & 2/2/2 & 2/2/2/2 \\
    2 & 3 & 3/3 & 3/3/3 & 3/3/3/3 \\
    3 & 4 & 4/4 & 4/4/4 & 4/4/4/4 \\
    4 & 5 & 5/5 & 5/5/5 & 5/5/5/5 \\
    5 & 6 & 6/6 & 6/6/6 & 6/6/6/6 \\
    6 & 13 & 13/15 & 13/15/17 & 13/15/17/19 \\
    7 & 17 & 17/19 & 19/21/23 & 21/23/25/27 \\
    8 & 21 & 21/23 & 25/27/29 & 29/31/33/35 \\
    % 9 & 25/-/-/- & 25/27/-/- & 33/31/35/- & 37/39/41/43 \\
    % 10 & 29/-/-/- & 29/31/-/-/ & 39/37/41/- & 45/47/49/51 \\
    $i>8$ & $(i-1) + 4$  & $(i-1) + 4$ &  $(i-1) + 6$  &   $(i-1) + 8$ \\
    \hline
    \end{tabular}
    \caption{Cycle in which each memory instruction is issued. Each cell stores the cycle for all active sub-cores sorted by cycles in ascending order.}
    \label{tab:memoryconsecutive}
    \vskip -0.8cm
\end{table}

\par
As regards latencies, we measured two different latencies for each instruction type when the instruction hits in the cache and there is a single thread in execution. The first latency is the elapsed time since a load is issued until the earliest time that a consumer or an instruction that overwrites the same destination register can issue. We refer to this as RAW/WAW latency (note that stores cannot generate register RAW/WAW dependencies). The second one is the elapsed time since a load or store is issued until the earliest time that an instruction that writes in a source register of the load/store can be issued. We refer to this time as WAR latency. The results are shown in \autoref{tab:mem_latencies}.

\begin{table}
    \centering
    \small
    \begin{tabular}{|l|c|c c|}
    \hline
    \multirow{3}{*}{\shortstack{Instruction}} & \multirow{3}{*}{\shortstack{Memory \\ address \\ register type}} & \multicolumn{2}{c|}{\multirow{2}{*}{\shortstack{Dependency type \\ latencies}}} \\
    & & & \\
    \cline{3-4}
    & & WAR & RAW/WAW \\
    \hline
    Load Global 32 bit & Uniform & 9 & 29 \\
    Load Global 64 bit & Uniform & 9 & 31 \\
    Load Global 128 bit & Uniform & 9 & 35 \\
    \hline
    Load Global 32 bit & Regular & 11 & 32 \\
    Load Global 64 bit & Regular & 11 & 34 \\
    Load Global 128 bit & Regular & 11 & 38 \\
    \hline
    Store Global 32 bit & Uniform & 10 & - \\
    Store Global 64 bit & Uniform & 12$^{*}$ & - \\
    Store Global 128 bit & Uniform & 16$^{*}$ & - \\
    \hline
    Store Global 32 bit & Regular & 14 & - \\
    Store Global 64 bit & Regular & 16 & - \\
    Store Global 128 bit & Regular & 20 & - \\
    \hline
    Load Shared 32 bit & Uniform & 9 & 23 \\
    Load Shared 64 bit & Uniform & 9 & 23 \\
    Load Shared 128 bit & Uniform & 9 & 25 \\
    \hline
    Load Shared 32 bit & Regular & 9 & 24 \\
    Load Shared 64 bit & Regular & 9 & 24 \\
    Load Shared 128 bit & Regular & 9 & 26 \\
    \hline
    Store Shared 32 bit & Uniform & 10 & - \\
    Store Shared 64 bit & Uniform & 12 & - \\
    Store Shared 128 bit & Uniform & 16 & - \\
    \hline
    Store Shared 32 bit & Regular & 12 & - \\
    Store Shared 64 bit & Regular & 14 & - \\
    Store Shared 128 bit & Regular & 18 & - \\
    \hline
    Load Constant 32 bit & Immediate & 10 & 26 \\
    Load Constant 32 bit & Regular & 29 & 29 \\
    Load Constant 64 bit & Regular & 29 & 29 \\
    \hline
    \texttt{LDGSTS} 32 bit & Regular & 13 & 39 \\
    \texttt{LDGSTS} 64 bit & Regular & 13 & 39 \\
    \texttt{LDGSTS} 128 bit & Regular & 13 & 39 \\
    \hline
    \end{tabular}
    \caption{Memory instructions latencies in cycles. Values with $^{*}$ are approximations as we were unable to gather these data.}
    \label{tab:mem_latencies}
    \vskip -0.9cm
\end{table}

\par
We can see that global memory accesses are faster if instructions use uniform registers for computing their addresses rather than regular registers. This difference is due to a faster address calculation. When using uniform registers, all threads in a warp share the same register, and thus, a single memory address needs to be computed. On the other hand, when using regular registers, each thread needs to compute a potentially different memory address.

\par
We can also observe that the latency of shared memory loads is lower than that of global memory. We can also observe that their WAR latency is the same for regular and uniform registers, whereas their RAW/WAW latency is one cycle lower for uniform registers. The fact that WAR latencies are equal for regular and uniform registers suggests that the address calculation for shared memory is done in the shared-structures, rather than in the local sub-core structures, so the WAR dependence is released once the source registers are read. 

\par
As expected, latencies also depend on the size of the read/written values. For WAR dependencies, the latency of loads does not change since the source operands are only for address calculation and thus, they are always the same size no matter the size of the loaded value. For store instructions, WAR latencies increase with the size of the written value to memory since this value is a source operand that also needs to be read from the register file. For RAW/WAW dependencies (only apply to loads), latencies increase as we increase the size of the read value, since more data needs to be transferred from the memory to the register file. We have measured that the bandwidth for this transfer is 512 bits per cycle.

\par
In addition, we can observe that the constant cache's WAR latency is significantly greater than loads to the global memory, whereas RAW/WAW latencies are a bit lower. We could not confirm any hypothesis that explains this observation. However, we discovered that accesses to the constant memory done by fixed-latency instructions go to a different cache level than load constant instructions. We confirmed this by preloading a given address into the constant cache through an \texttt{LDC} instruction and waiting for the instruction until it is complete. Then, we issued a fixed-latency instruction using the same constant memory address and measured that the issue was delayed 79 cycles, which corresponds to a miss, instead of causing no delay, which would correspond to a hit. Therefore, fixed-latency instructions accessing the constant address space use the L0 FL (fixed latency) constant cache, while accesses made through \texttt{LDC} instructions utilize the L0 VL (variable latency) constant cache.

\par
Finally, we analyze \texttt{LDGSTS} instruction, which was introduced to reduce the pressure of the register file and improve the efficiency of transferring data to the GPU \cite{patentLDGSTS}. It loads data from global memory and stores it directly into the shared memory without going through the register file, which saves instructions and registers. We can see that its latency is the same regardless of the granularity of the instruction. WAR dependencies have the same latency for all granularities since they are released when the address is computed. The RAW/WAR dependency is released when the instruction's read step has finished, regardless of the instruction's granularity.

\section{Modeling}\label{sec:modeling}

\par
We have designed from scratch the SM/core model of the Accel-sim framework simulator~\cite{accelsim} by modifying the pipeline to implement all the details explained in~\autoref{sec:modernGPUArch},~\autoref{sec:design} and depicted in~\autoref{fig:ourCore}. The main new components are outlined below.

\par
First, we have added an L0 instruction cache with a stream buffer prefetcher for each sub-core. L0 instruction and constant caches are connected with a parameterized latency to an L1 instruction/constant cache. We have chosen the size of caches, hierarchy, and latencies according to previous measurements done or the Ampere architecture described by Jia et al~\cite{dissectingAmpere}.

\par
We have modified the Issue stage to support the control bits, the tag look-up to the newly added L0 constant caches for fixed-latency instructions, and the new CGGTY issue scheduler. We included the control stage, in which instructions increase the dependence counters, and the Allocate stage, in which fixed-latency instructions check for conflicts in the access to the register file and the register file cache. 

\par
Regarding memory instructions, we have modeled a new unit per sub-core and a unit shared among sub-cores, with the latencies presented in the previous section. 

\par
Additionally, as Abdelkhalik et al.~\cite{demystifyingAmpere} have demonstrated that the latency of a tensor core instruction depends on its operand's numeric types and sizes, so we have adapted the model to use the correct latency for each operand type and size.

\par
Other details that we have modeled are the execution pipeline shared by all sub-cores for double precision instructions in architectures without dedicated double precision execution units in each sub-core. Moreover, we accurately model the timing for reading/writing operands that use multiple registers, which was previously approximated by using just one register per operand. Furthermore, we fixed some inaccuracies in instruction addresses reported in a previous work~\cite{cams2023paper}.

\par
Apart from implementing the new SM/core model in the simulator, we have extended the tracer tool. The tool has been extended to dump the ID of all types of operands (regular registers, uniform registers, predication registers, immediate, etc.). Another important extension is the capability to obtain the control bits of all instructions since NVBit does not provide access to them. This is done by obtaining the SASS through the CUDA binary utilities~\cite{cudaBinaryUtils} at compile time. This implies modifying the compilation of the applications to generate microarchitecture-dependent code at compile time instead of using a just-in-time compilation approach. Unfortunately, for a few kernels (all of them belonging to Deepbench), NVIDIA tools do not provide the SASS code, which prevents obtaining the control bits for these instructions. To simulate these applications, we use a hybrid mode for dependencies where traditional scoreboards are employed in kernels that do not have the SASS code; otherwise, control bits are used. 

\par
We have extended the tool to capture memory accesses to the constant cache or global memory through descriptors. Despite the latter type of accesses being claimed to be introduced in Hopper~\cite{cudaBinaryUtils}, we have noticed that Ampere already uses them. Descriptors are a new way of encoding memory references that uses two operands. The first operand is a uniform register to encode the semantics of the memory instruction, while the second encodes the address. We have extended the tracer to capture the address. Unfortunately, the behavior encoded in the uniform register remains untracked.

\par
We plan to make public all the simulator and tracer changes made in the Accel-sim framework.

%No se si habria que decir que soporta mas aplicaciones que el simulador original. Pero que por ser justos nos comparamos solo con las que han acabado en ambos
\section{Validation}\label{sec:valiadation}

\par
In this section, we evaluate the accuracy of our proposed microarchitecture for the GPU cores. First, we describe the methodology that we have followed in \autoref{subsec:methodoloy}. Then, we validate the design in \autoref{subsec:cycle_accuracy}. Next, we examine the impact of the register file cache and the number of register file read ports on accuracy and performance in \autoref{subsec:eval_rf}. Later, we study the impact of two different components in the design, such as the instruction prefetcher in \autoref{subsec:results_instruction_prefetching} and an analysis of the dependence checking mechanisms in \autoref{subsec:deps_analysis}. Finally, we discuss how the model can be seamlessly adapted for NVIDIA architectures other than Ampere in \autoref{subsec:use_other_archs}.

\subsection{Methodology}\label{subsec:methodoloy}

\par
We validate the accuracy of the proposed GPU core by comparing the results of our version of the simulator against hardware counter-metrics obtained in a real GPU. We use \numberOfGPUsWord{} different Ampere \cite{amperePaper} GPUs whose specifications are shown in \autoref{tab:gpus_specs_and_results}. All the GPUs use CUDA 11.4 and NVBit 1.5.5. We also compare our model/simulator with the vanilla Accel-sim simulator framework, since our model is built on top of it.

\par
We use a wide variety of benchmarks from \numberOfBenchSuites{} different suites. A list of the suites that have been employed and the number of applications and different input data sets can be found in \autoref{tab:bench_suites}. In total, we use \numberOfDiffCombinations{} benchmarks, \numberOfDiffApplications{} of those being different applications, and the rest corresponding to just changing the input parameters.

%\par
%We profiled each benchmark to gather the number of execution cycles in the real GPU. Later, we compare the real hardware execution cycles against the execution cycles reported in each simulation to compute the cycle error. Finally, we calculate the mean absolute error (MAE) of cycles among all the benchmarks for each simulator. Moreover, we employ the Pearson correlation coefficient to measure the simulators' robustness.

\begin{table}
    \centering
    \small
    \begin{tabular}{ |c|c|c|  }
    \hline
    Suite & Applications & Input sets \\
    \hline
    Cutlass \cite{cutlass} & 1 & 17 \\
    Deepbench \cite{deepbenchWeb} & 3 & 27 \\
    Dragon \cite{dragon} & 4 & 6 \\
    GPU Microbenchmark \cite{accelsim} & 15 & 15 \\
    ISPASS 2009 \cite{gpgpusim3} & 8 & 8 \\
    Lonestargpu \cite{lonestar} & 3 & 6 \\
    Pannotia \cite{pannotia} & 7 & 11 \\
    Parboil \cite{parboil} & 6 & 6 \\
    Polybench \cite{polybench} & 8 & 8 \\
    Proxy Apps DOE \cite{proxy} & 3 & 4 \\
    Rodinia 2 \cite{rodinia} & 10 & 10 \\
    Rodinia 3 \cite{rodinia} & 15 & 25 \\
    \hline
    Total & \numberOfDiffApplications{} & \numberOfDiffCombinations{} \\
    \hline
    \end{tabular}
    \caption{Benchmarks suites.}
    \label{tab:bench_suites}
    \vskip -0.9cm
\end{table}

%Unfortunately, we are unable to test our findings in Ada \cite{adaPaper} and Hopper \cite{hopperPaper} due to the incompatibility of NVBit \cite{NVBIT} with the newest CUDA versions and architectures.

\subsection{Performance Accuracy}\label{subsec:cycle_accuracy}

\begin{table*}
    \centering
    \small
    \begin{tabular}{ |c|c|c|c|c||c|  }
    \cline{2-6}
    \multicolumn{1}{c|}{} & \multicolumn{4}{c||}{Ampere} & \multicolumn{1}{c|}{Turing} \\
    \cline{2-6}
    \multicolumn{1}{c|}{} & RTX 3080 & RTX 3080 Ti & RTX 3090 & RTX A6000 & RTX 2080 Ti \\
    \hline
    \multicolumn{6}{|c|}{Specifications} \\
    \hline
    Core Clock & 1710 MHz & 1365 MHz & 1395 MHz & 1800 MHz & 1350 MHz \\
    Mem. Clock  & 9500 MHz & 9500 MHz & 9750 MHz & 8000 MHz & 7000 MHz\\
    \# SM & 68 & 80 & 82 & 84 & 68 \\
    \# Warps per SM & 48 & 48 & 48 & 48 & 32 \\
    \shortstack{Total Shared\\mem./L1D per SM} & \shortstack{128 KB} & \shortstack{128 KB} & \shortstack{128 KB} & \shortstack{128 KB} & \shortstack{96 KB} \\
    \# Mem. part. & 20 & 24 & 24 & 24  & 22 \\
    Total L2 cache & 5 MB & 6 MB & 6 MB & 6 MB & 5.5 MB \\
    \hline
    \multicolumn{6}{|c|}{Validation} \\
    \hline
    Our model MAPE & \RESEVAL{MaeRTX3080Us} & \RESEVAL{MaeRTX3080TiUs} & \RESEVAL{MaeRTX3090Us} & \RESEVAL{MaeRTXA6000Us} & \RESEVAL{MaeRTX2080TiUs} \\
    Accel-sim MAPE & \RESEVAL{MaeRTX3080Accel} & \RESEVAL{MaeRTX3080TiAccel} & \RESEVAL{MaeRTX3090Accel} & \RESEVAL{MaeRTXA6000Accel} & \RESEVAL{MaeRTX2080TiAccel} \\
    Our model Correl. & \RESEVAL{CorrelRTX3080Us} & \RESEVAL{CorrelRTX3080TiUs} & \RESEVAL{CorrelRTX3090Us} & \RESEVAL{CorrelRTXA6000Us} & \RESEVAL{CorrelRTX2080TiUs} \\
    Accel-sim Correl. & \RESEVAL{CorrelRTX3080Accel} & \RESEVAL{CorrelRTX3080TiAccel} & \RESEVAL{CorrelRTX3090Accel} & \RESEVAL{CorrelRTXA6000Accel} & \RESEVAL{CorrelRTX2080TiAccel} \\
    \hline
    \end{tabular}
    \caption{GPUs specifications and performance accuracy.}
    \label{tab:gpus_specs_and_results}
    \vskip -0.4cm
\end{table*}

\par
\autoref{tab:gpus_specs_and_results} shows the reported mean percentage absolute error (MAPE) for both models (our model and Accel-sim) with respect to the real hardware for each of the GPUs. We can see that our model is significantly more accurate than Accel-sim in all the evaluated GPUs, and for the biggest GPU, the \targetGPUAmpere{}, the MAPE is less than half that of Accel-sim. Regarding the correlation, both models are very similar, but our model is slightly better.

\begin{figure}[ht]
  \centering
  \includegraphics[trim={0.0cm 0.0cm 0.0cm 0.0cm},clip,width=8.5cm]{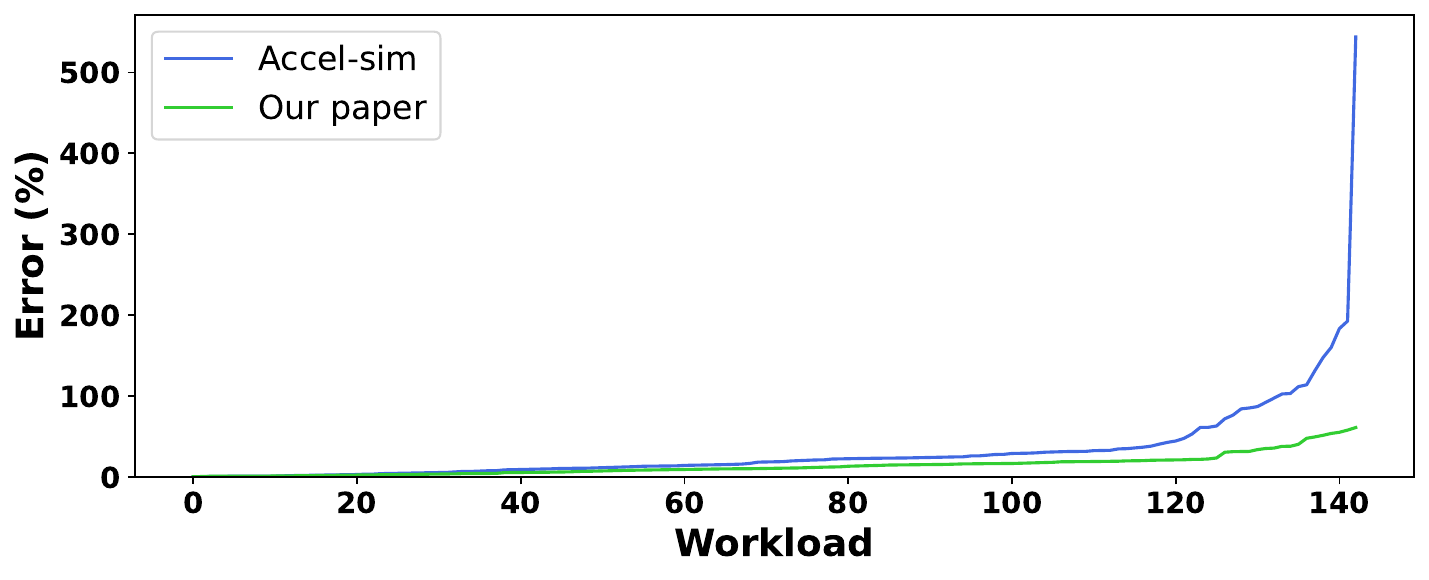}
  \vskip -0.5cm
  \caption{\targetGPUAmpere{} percentage absolute error for each benchmark. Benchmarks are sorted by error in ascending order in each of the configurations.}
  \label{fig:cycles_results}
  \Description{Figure that compares in an accumulative way Accel-sim and our model for the NVIDIA RTX A6000} % Useless, used to remove warnings
  \vskip -0.5cm
\end{figure}

\par
\autoref{fig:cycles_results} shows the APE of both models for the \targetGPUAmpere{} and each of the \numberOfDiffCombinations{} benchmarks sorted in increasing error for each of the models. We can see that our model consistently has less absolute percentage error than Accel-sim for all applications, and the difference is quite significant for half of the applications. Moreover, we can observe that Accel-sim has an absolute percentage error greater or equal to $100\%$ for \RESEVAL{AeMoreThan100ErrorAccel} applications, and it reaches \RESEVAL{AeBiggestErrorAccel} in the worst case, whereas our model never has an absolute percentage error greater than \RESEVAL{AeBiggestErrorUs}. If we look at the 90th percentile as an indication of the tail accuracy, Accel-sim absolute percentage error is \RESEVAL{MaeQ90Accel}, whereas it is \RESEVAL{MaeQ90Us} for our model. This proves that our model is significantly more accurate and robust than the Accel-sim model.

%Si me da tiempo create una web anonima y pondre enlace donde see vean todos los resultados por workload y por configuración.
%Si da tiempo al sert, simulare Turing, see añadira a la tabla y tan solo con una frase de que que el model también funciona con otras arquitecturas será suficiente.

%This aligns with quartile 75, where Accel-sim has a \RESEVAL{MaeQ75Accel} error while \RESEVAL{MaeQ75Us} in our model, demonstrating that our model is more accurate and robust for a vast majority of applications.

\subsection{Sensitivity Analysis of Instruction Prefetching}\label{subsec:results_instruction_prefetching}

\par
The characteristics of the stream buffer instruction prefetcher have high impact on the global model accuracy. In this section, we analyze the error of different configurations, including disabling the prefetcher, having a perfect instruction cache, and a stream buffer prefetcher with sizes 1, 2, 4, 8, 16, and 32 entries. All the configurations are based on the \targetGPUAmpere{}. The MAPE for each configuration is shown in~\autoref{tab:prefetching_analysis}. We can see that the best accuracy is obtained with a stream buffer of size \bestStreamBufferSizeNumber{}. 

\begin{table*}
    \centering
    \small
    \begin{tabular}{ |c|c|c|c|c|c|c|c|c|  }
    \hline
    & Disabled & 1 & 2 & 4 & 8 & 16 & 32 & Perfect ICache \\
    \hline
    MAPE & 56.61\% & 43.94\% & 28.59\% & 18.55\% & 14.67\% & 13.98\% & 14.35\% & 15.2\% \\
    \hline
    Speed-up & 1 & 1.08x & 1.2x & 1.33x & 1.42x & 1.47x & 1.46x & 1.58x \\
    \hline
    \end{tabular}
    \caption{MAPE of different prefetcher configurations in the \targetGPUAmpere{} and speed-up of each configuration with respect to prefetching disabled.}
    \label{tab:prefetching_analysis}
    \vskip -0.4cm
\end{table*}

\par
We can draw an additional conclusion from the speed-up results shown in \autoref{tab:prefetching_analysis}: a straightforward prefetcher, such as a stream buffer, behaves close to a perfect instruction cache in GPUs. This is because the different warps in each sub-core usually execute the same code region and the code of typical GPGPUs applications do not have a complex control flow, so prefetching $N$ subsequent lines usually performs well. Note that since GPUs do not predict branches, it is not worth implementing a Fetch Directed Instruction prefetcher \cite{fdpPrefetching} because it would require the addition of a branch predictor.

\par
Regarding simulation accuracy, we conclude that when instruction cache enhancements are not under investigation, using a perfect instruction cache usually yields comparable accuracy with faster simulation speeds. However, for benchmarks where control flow is relevant —such as \textit{dwt2d}~\cite{rodinia}, \textit{lud}~\cite{rodinia}, or \textit{nw}~\cite{rodinia}— employing a perfect instruction cache or omitting stream buffers results in significant inaccuracies (more than 20\% difference compared to a perfect instruction cache and more than 200\% respect to an instruction cache without prefetching). This inaccuracy arises because a perfect instruction cache fails to capture the performance penalties incurred by frequent jumps between different code segments, while the absence of prefetching overly penalizes the execution of other parts of the program, which also reveals that there is an opportunity for improvement in those benchmarks.

%\vskip -1cm
\subsection{Sensitivity Analysis of the Register File Architecture}\label{subsec:eval_rf}

\par
~\autoref{tab:rf_analysis} illustrates how the presence of a register file cache and an increased number of register file read ports per bank affect simulation accuracy and performance. It also shows the results for an Ideal scenario where all operands can be retrieved in a single cycle. The average performance and accuracy across all benchmarks are similar for all configurations. However, a closer examination of specific benchmarks —such as the compute-bound \textit{MaxFlops} (Accel-sim GPU Microbenchmark)~\cite{accelsim} and Cutlass~\cite{cutlass} configured with the \textit{sgemm} parameter —reveals more nuanced behavior. Both benchmarks rely heavily on fixed-latency arithmetic instructions, which are particularly sensitive to stalls caused by register file access limitations because they typically use three operands per instruction.

\begin{table}
    \centering
    \small
    \begin{tabular}{ |c|c|c|c|c|  }
    \hline
    & 1R RFC on & 1R RFC off & 2R RFC off & Ideal \\
    \hline
    MAPE & \RESEVAL{MaeRTXA6000Us} & 16.05\% & 13.38\% & 13.57\%  \\
    Speed-up & 1x & 0.984x & 1.012x & 1.013x \\
    \hline
    MaxFlops APE & 2.82\% & 2.82\% & 28.97\% & 28.97\% \\
    MaxFlops speed-up & 1x & 1x & 1.44x & 1.44x \\
    \hline
    Cutlass APE & 9.72\% & 39.35\% & 0.97\% & 2.3\% \\
    Cutlass speed-up & 1x & 0.79x & 1.11x & 1.12x \\
    \hline
    \end{tabular}
    \caption{MAPE of different RF configurations in the \targetGPUAmpere{} and speed-up of each configuration with respect to the baseline (1 read port and RFC enabled).}
    \label{tab:rf_analysis}
    \vskip -0.7cm
\end{table}

\par
For \textit{MaxFlops}, performance is identical regardless of whether the RFC is present, since only one static instruction makes use of it. Notably, performance improves dramatically by approximately 44\% when two read ports per register file bank are employed. This improvement is logical given that three operands per instruction are common and four read ports (two per bank) are sufficient to meet demand. In contrast, from a simulation accuracy perspective, the configuration of the two read ports exhibits a significant deviation.

\par
In the case of Cutlass with \textit{sgemm}, a single-port configuration without a register file cache leads to a substantial performance degradation (0.78x). This performance drop is consistent with the observation that 35.9\% of the static instructions in the program make use of the register file cache in at least one of their operands. However, introducing a register file with two read ports per bank yields a 12\% performance improvement, which suggests that there is room for improvement in the organization of the register file and its cache.

\par
In summary, the register file architecture, including its cache, has an important effect on individual benchmarks, so its accurate model is important. A single port per bank plus a simple cache performs close to a register file with an unbounded number of ports on average, but for some individual benchmarks the gap is significant, which suggests that this may be an interesting area for research.

%when the compiler effectively mitigates register conflicts through optimal operand bank mapping and register file cache usage, the resulting performance is only about 13\% below ideal. Moreover, a two-read-port configuration achieves nearly ideal performance. However, it is important to consider that the register file in GPU is huge, and increasing the number of ports comes at a significant cost, as area scales quadratically\mbox{~\cite{areaQuadraticRF}} while energy scales linearly\hbox{~\cite{energyLinearRF}}.}

%\vspace{-2cm}
\subsection{Analyzing Dependence Management Mechanisms}\label{subsec:deps_analysis}

\par
In this subsection, we analyze the impact on performance and the area of the software-hardware dependence handling mechanism explained in this paper and compare them with the traditional scoreboard method used by former GPUs. \autoref{tab:dependence_checking_mechanisms} shows the results for both metrics. Area overhead is reported relative to the area of the regular register file of an SM, which is 256 KB.  

\begin{table}
    \centering
    \small
    \begin{tabular}{ c c|c|c|  }
    \cline{3-4}
    & & \multicolumn{2}{c|}{\shortstack{Scoreboarding with different \\ number of maximum consumers}} \\
    %& & \multicolumn{2}{c|}{Scoreboarding with different number of maximum consumers} \\
    \cline{2-4}
    \multirow{1}{*}{} & \multicolumn{1}{|c|}{\multirow{1}{*}{Control bits}} & \multirow{1}{*}{\shortstack{63}} & \multirow{1}{*}{\shortstack{Unlimited}} \\
     % & \multicolumn{1}{|c|}{} & & & \\
     % & \multicolumn{1}{|c|}{} & & & \\
     % & \multicolumn{1}{|c|}{} & & &\\
    \hline
    \multicolumn{1}{|c|}{Speed-up} & 1 & 0.97x & 0.97x \\
    \hline
    \multicolumn{1}{|c|}{Area overhead} & $0.09\%$ & $5.32\%$ & - \\
    \hline
    \multicolumn{1}{|c|}{MAPE} & $13.98\%$ & $14.87\%$ & $14.87\%$ \\
    \hline
    \end{tabular}
    \caption{Speed-up, area overhead and MAPE of different dependence management mechanisms.}
    \label{tab:dependence_checking_mechanisms}
    \vskip -1.2cm
\end{table}

\par
%Regarding area, we first compute the storage size required for each mechanism for a single warp. Then, we scale it to the amount required in a single SM by multiplying the storage required by a single warp by the total number of warps in an SM (48 in Ampere analyzed GPUs). Finally, we compute the overhead relative to the storage of the regular register file of an SM, which is 256 KB.

\par
A mechanism based on the traditional scoreboard requires as many entries as registers that can be written, that is,  332 entries per warp (255 for the regular registers, 63 for the uniform registers, 7 for the predicate registers, and 7 for the uniform predicate registers). Additionally, two scoreboards are needed: one for WAW/RAW hazards and another for WAR hazards \cite{warHazards}, because even though issue is in-order, the read-write of operands may occur out-of-order due to variable-latency instructions. WAR hazards may occur because variable-latency instructions, such as memory instructions, are queued after being issued and may read their source operands after a younger arithmetic instruction writes its result. While the first scoreboard requires only a single bit per entry, the second one requires more hardware as the number of consumers per entry increases. Assuming support for up to 63 consumers per entry, a single warp would require 2324 bits ($332 + 332 \times log2(63+1)$) for dependency handling. For the entire SM, this translates to 111,552 bits, which is $5.32\%$ of the register file size.

\par 
In contrast, the software-hardware mechanism presented in this paper requires six \textit{Dependence Counters} of six bits each, a \textit{Stall Counter} of four bits, and a yield bit. This amounts to just 41 bits per warp or 1968 bits per SM. In terms of overhead, this is only $0.09\%$ of the register file size, which is much less than the scoreboard alternative.

%\par 
%Examining the performance obtained with the scoreboard mechanism, we observe that when the WAR scoreboard supports up to one or three consumers performance is significantly degraded in comparison to the control bits alternative. As the number of consumers increases, performance improves. We have evaluated configurations of up to 63 consumers per entry in the scoreboard because it is the maximum value supported by a \textit{Dependence counter} of control bits. Notably, control bits outperform all the scoreboard scenarios, demonstrating that this design is an excellent cost-effective approach for dependence handling.

%Notably, these last three scoreboard scenarios achieve performance similar to the control bits mechanism, confirming control bits as an excellent cost-effective approach for dependence handling.

\par 
In summary, the software-hardware codesign based on control bits outperforms other alternatives and it has a negligible area overhead. The difference in overhead with respect to scoreboarding becomes even more significant in GPUs that support up to 64 warps per SM, such as NVIDIA Hopper \cite{hopperPaper}, which has an overhead of $0.13\%$ for the control bits alternative and $7.09\%$ for the scoreboard mechanism with 63 consumers. An additional insight is that using a scoreboard mechanism composed of two scoreboards (one for RAW/WAW and another one for WAR hazards with up to 63 consumers) is comparable in simulation accuracy to using control bits. Therefore, it is a valid alternative for applications that do not expose the values of control bits, such as some kernels of the Deepbench benchmark suite.

\subsection{Portability to other NVIDIA architectures}\label{subsec:use_other_archs}

\par
In this paper, we focus on the NVIDIA Ampere architecture. However, the findings we have exposed in the paper apply to other architectures, such as Turing. In addition to reporting results for four Ampere GPUs, \autoref{tab:gpus_specs_and_results} presents results for a Turing GPU, showing a \RESEVAL{MaeBiggestImprovementTuring} MAPE improvement over Accel-sim for the \targetGPUTuring{}.

\par
Although we have validated the model on Turing and Ampere, we believe that our findings remain applicable to other NVIDIA architectures. NVIDIA’s public announcements and SM diagrams indicate that significant architectural changes have been limited to enhancements in Tensor cores, ray-tracing units, and minor features such as distributed shared memory between SMs of the same TPC~\cite{hopperPaper}. Nonetheless, estimating the latencies for some instructions, such as memory, will be necessary to adapt the model to these other architectures.

\section{Related Work}\label{sec:relatedwork}

\balance % Remove warning

\par
Simulators are the primary tool for evaluating ideas in computer architecture in academia and industry \cite{SimSurvey} due to the low cost of assessing new designs. Moreover, they are a great tool to check if designs are close to actual hardware. GPGPUs are no exception, and a leading vendor such as NVIDIA has exposed part of the process of creating their in-house simulator, NVIDIA Architectural Simulator (NVArchSim or NVAS) \cite{NVAS}. In the academic sphere, there are two popular public-domain simulators. The first one is MGPUSim \cite{mgpusim}, which models the AMD GCN 3 architecture and targets multi-GPU systems supporting virtual memory. An alternative simulator is the Accel-Sim framework \cite{accelsim}, a cycle-accurate state-of-the-art trace-driven simulator supporting CUDA applications that models NVIDIA-like modern architectures, which it is based on the former GPGPU-Sim 3 simulator \cite{gpgpusim3}.

\par
In the literature, there are different works that try to reverse engineering CPU architectural components, such as the Intel Branch Predictor~\cite{intelBranchPredictor} or Intel cache designs~\cite{intelReverseCacheAddress, intelReverseCacheSlice} to name just a few. 

\par
Regarding NVIDIA GPUs, several works have been pursued to unveil specific components of GPUs. Ahn et al.~\cite{reverseNoc} and Jin et al.~\cite{UncoveringNOC} target to reverse engineer the Network on Chip of the NVIDIA Volta and Ampere architectures, respectively. Lashgar et al.~\cite{reverseMshrPRT} study NVIDIA Fermi's and Kepler's capabilities for handling memory requests. Jia et al.~\cite{dissectingVolta, dissectingTuring} present some cache features such as line size and associativity, the latency of instructions, and some details of the register file of Volta and Turing. Khairy et al.~\cite{voltaCaches} explore the L1 data cache and L2 cache designs of Volta. Abdelkhalik et al.~\cite{demystifyingAmpere} establish the relation between PTX and SASS instructions of Ampere and its execution latency. Regarding tensor cores, different works~\cite{dissectingVolta, dissecTensorCoresVolta1, dissecTensorCoresVolta2, dissectingTuring, dissecTensorCoresTuringVolta1, dissecTensorCoresTuringVolta2, dissecTensorCoresAmpere1, dissecTensorCoresAmpere2} have investigated their programmability and micro-architecture. TLBs of Turing and Ampere have been reverse-engineered by Zhang et al. \cite{tunnelsTLB} with the purpose of realizing an attack on Multi-Instance GPUs. Concerning the control flow of modern NVIDIA GPU architectures, Shoushtary et al.~\cite{mojtabaControlFlow} define a plausible semantic for control flow instructions of Turing with just a $1.03\%$ error with respect to real hardware traces. The GPU task scheduler and the interaction with an ARM CPU in the NVIDIA Jetson TX2 are studied by Amert et al.~\cite{reverseTX2}. Finally, Wong et al.~\cite{demystifyingGPU} describe many components of the early Tesla architecture, such as caches, TLB, SIMT control flow behavior, and Cooperative Thread Array (CTA) barriers.

\par
Regarding previous works that focused on other GPU vendors, Gutierrez et al.~\cite{amdGap1} show that directly employing the AMD GPU machine ISA rather than intermediate languages is crucial for accurately assessing bottlenecks and potential solutions during simulation. Moreover, the GAP tool~\cite{amdGap2} identifies discrepancies between real AMD GPUs and their gem5~\cite{gem5} simulated counterparts, which has led to improvements in gem5's AMD GPU accuracy~\cite{amdGap3,amdGap4}. Finally, Gera et al.~\cite{intelGPUs} introduce a simulation framework and characterize Intel’s integrated GPUs.

\par
Compiler hints (a.k.a. control bits) have been used in GPUs at least since the NVIDIA Kepler architecture~\cite{KeplerASPaper, KeplerASRepo}. A description of these control bits was published by Gray et al.~\cite{maxas}. In Kepler, Maxwell, and Pascal architectures, one out of each 3 to 7 instructions is usually a compiler-inserted hint instruction. Jia et al.~\cite{dissectingVolta, dissectingTuring} revealed that newer architectures such as Volta or Turing have increased the instruction bit size from 64 to 128 bits. As a result, newer architectures have shifted from having specific instructions for hints to including the hint bits in each instruction. These bits are intended not only to improve the hardware performance but also to ensure the program's correctness by preventing data hazards. CPUs have used compiler hints to help hardware make better decisions about how to use the different resources~\cite{HardwareAwareCompilation}. These hints have been used to support different components of the CPUs, such as data bypass, branch prediction, and caches, among others.

\par
To the best of our knowledge, our work is the first one to unveil the GPU core microarchitecture of modern NVIDIA GPUs and develop an accurate microarchitecture simulation model. Some novel features discovered in our work are: a complete definition of the semantics of the control bits and the microarchitectural changes to support them, the behavior of the issue scheduler, the microarchitecture of the register file and its associated cache, and various aspects of the memory pipeline. These aspects are critical for an accurate modeling of modern NVIDIA GPUs.

\section{Conclusion}\label{sec:conclusion}

\par
This paper unveils the industry's modern NVIDIA GPU microarchitecture by reverse engineering it on real hardware. We dissect the issue stage logic, including analyzing warp readiness conditions and discovering that the issue scheduler among warps follows a CGGTY policy. In addition, we unveil different details of the register file, such as the number of ports and their width. Also, we reveal how the register file cache works. Moreover, this paper exhibits some important characteristics of the memory pipeline, like the size of load/store queues, contention between sub-cores, and how latencies are affected by memory instruction granularity accesses. Furthermore, we analyze the fetch stage and propose one that meets the requirements of modern NVIDIA GPUs.

\par
Additionally, the paper compendiums previous public information about control bits by organizing, explaining it in detail, and extending it.

\par
In addition, we model all these details in a simulator and compare this new model against real hardware, demonstrating that it is closer to reality than the previous models by improving its accuracy in cycles by more than \RESEVAL{MaeBiggestImprovement}.

\par
Besides, we demonstrate that instruction prefetching with a simple stream buffer in GPUs performs well in terms of simulation accuracy and performance, approaching a perfect instruction cache. Also, we show how the dependence management mechanism based on control bits used in modern NVIDIA GPUs outperforms other alternatives, such as traditional scoreboarding.

\par
Finally, we investigate how the register file cache and the number register file read ports affect simulation accuracy and performance.

\par
Overall, we can conclude that GPUs are hardware-compiler codesign where the compiler guides the hardware in handling dependencies and introduces hints that can improve performance and energy.

% Print the bibliography
\bibliographystyle{ACM-Reference-Format}
\bibliography{refs}

\end{document}